# Unconventional superconductivity emerging along with the strange-metal behavior in UAs$_2$ under pressure

Qing Li[1†], Zhe-Ning Xiang[1†], Bin-Bin Zhang[2†], Ying-Jie Zhang[1†], Chaofan Zhang[2*], Hai-Hu Wen[1*]

[1] National Laboratory of Solid State Microstructures and Department of Physics, Collaborative Innovation Center of Advanced Microstructures, Nanjing University, Nanjing 210093, China.

[2] College of Advanced Interdisciplinary Studies & Nanhu Laser Laboratory, National University of Defense Technology, Changsha 410073, China.

*Corresponding authors: C.Zhang@nudt.edu.cn; hhwen@nju.edu.cn

**Abstract:**

The recently discovered spin-triplet superconductor candidate UTe$_2$ with $T_c \sim 2$ K [1,2] has attracted enormous attention because it possesses many interesting properties, such as the extremely high upper critical field $H_{c2}(0)$ [1-3], chiral superconductivity [4] and spontaneous time-reversal symmetry breaking [5], *etc.*, all these suggest that it may be the long-sought spin-triplet superconductor. Here we report the discovery of superconductivity up to $T_c \approx 4$ K in one of its siblings, i.e., UAs$_2$ under high pressures. Interestingly, the UAs$_2$ shows metallic behavior with an antiferromagnetic (AFM) transition at about 274 K under ambient pressure. Upon applying pressure, this transition is pushed down to lower temperatures with improved electric conductivity. When the pressure rises to about 20-22 GPa, superconductivity occurs together with the emergence of a linear temperature dependence of normal state resistance, the latter is a hallmark of the strange-metal state. The superconductivity with the highest $T_c \approx 4$ K is reached under a pressure of about 26.8 GPa, and it is robust against magnetic field with the upper critical field $\mu_0 H_{c2}(0) \sim 12$ T, far beyond the Pauli limit. Higher pressures will suppress the superconductivity and bring back the Fermi liquid behavior, showing a clear signature of quantum criticality. Our results open a new avenue for investigating the unconventional superconductivity concerning the mysterious 5$f$-band electrons in this uranium-based system.

# Introduction

Unconventional superconductivity with spin-triplet pairing has attracted widespread interests in condensed matter physics because it offers fruitful new physics and potential applications for topological quantum computation [6, 7, 8]. However, intrinsic spin-triplet superconductors with odd parity rarely exist in nature. Just a few years ago, unconventional superconductivity was discovered in UTe$_2$ [1, 2]. Interests arise immediately since it locates at the paramagnetic end, and could be the member of a ferromagnetic superconductor series like UGe$_2$ [9], URhGe [10], and UCoGe [11]. The superconductivity in UTe$_2$ has been extensively explored within a short period in the quest for possible spin-triplet pairing, and many experimental results are very promising towards this direction. These interesting results include, for example, the extremely large upper critical fields and coexistence of multiple re-entrant superconducting phases under a high magnetic field [1,2, 3, 12], the very small change of Knight shift on entering the superconductive state [1], the evidence for the presence of chiral modes existing inside the superconducting gap [4], and so on. [13,14] Recently, scanning tunneling microscopy measurements reveal the existence of charge density wave and pair density wave in the superconducting states [15,16]. Interestingly, high pressure can generate a second superconducting phase with a maximum $T_c$ of about 3.0 K at 1.2 GPa. After exceeding 1.4 GPa, the possible antiferromagnetic ordered phase appears along with the suppression of superconductivity [17, 18]. Under higher pressures above 3.5-4 GPa, a structure phase transition from orthorhombic (*Immm*) to tetragonal (*I4/mmm*) occurs and a new superconducting phase at pressures higher than 7 GPa was detected [19].

Previous studies have reported the synthesis and investigations on a series of uranium di-pnictide U$X_2$ ($X$ = P, As, Sb, Bi) compounds, which crystallize in the tetragonal structure [20]. The typical crystal structure of UAs$_2$ is plotted in Fig. 1**a;** there are two types of As sites in UAs$_2$ marked as As$_1$ and As$_2$. All these compounds are antiferromagnets with relatively high Néel temperatures ($T_N$), and the UAs$_2$ has the highest value ($T_N$ = 273 K) among them [21, 22]. As evidenced from the neutron diffraction data [23, 24], the magnetic study reveals that the ferromagnetic sheets composed by U ions can form a stacking structure along *c*-axis with the spin orientation perpendicular to the basal plane and a configuration of up/down/down/up for *X* = P, As, Sb and up/down/up/down for X = Bi. The magnetic unit cell is doubled along the *c*-axis compared to the crystal structure in UAs$_2$ (Fig. 1b). Transport and quantum oscillation measurements reveal that the U*X*$_2$ series are metallic and have strong two-dimensional electronic character with cylindrical Fermi surface sheets [25, 26]. Furthermore, angle-resolved photoemission studies on USb$_2$ and UAs$_2$ have observed the gap opening below $T_N$ and the evidence of hybridization between the 5*f*-band electrons and conduction electrons [27, 28].

The strange-metal state is an extraordinary feature of unconventional superconductors, in which the electrical resistance shows a linear relationship with temperature in the low-temperature limit. Such a unique strange-metal behavior has been widely observed in different material systems with unconventional superconductivity, such as the cuprates [29], iron-based and nickel-based superconductors [30, 31], heavy Fermion system [32], and so on [33, 34, 35]. Usually, the quantum phase transitions connected with unconventional superconductivity and the strange-metal state exist in an extended phase area near the quantum critical point (QCP). Therefore, it is generally believed that the unconventional superconductivity, including spin triplet parity, is generated from the fluctuations associated with this quantum criticality [36].

Considering the possible spin-triplet pairing states with intriguing superconducting properties in U-based heavy fermion system and the dual character of the mysterious 5*f*-band electrons, i.e., itineracy and localization, it is essential to explore the unconventional superconductivity and investigate the interplay between magnetism and superconductivity in U-based materials. In this work, by applying high pressure on UAs$_2$, we successfully generated a superconducting state when the magnetic transition is suppressed. The sudden change of room temperature resistance at around 20 GPa may originate from a possible structure phase transition from tetragonal to orthorhombic symmetry as proposed in previous literatures [37, 38]. In the superconducting state, we found that the maximum transition temperature ($T_c$) is about 4.0 K with a linear temperature dependence of resistivity (strange metal behavior) in the normal state. More importantly, we found that the superconducting upper critical field [$\mu_0 H_{c2}(0)$] clearly exceeds the Pauli paramagnetic limit. These new results, including the high $\mu_0 H_{c2}(0)$ and the strange metal behavior in the normal state, indicate a possible exotic pairing mechanism in UAs$_2$ under high pressure.

## Results

**1. Sample characterization at ambient pressure**

UAs$_2$ crystallizes into a tetragonal anti-Cu$_2$Sb type structure with the space group *P*4/*nmm* (No. 129; $Z = 2$). The structure can be seen as composed of a basal-plane of As$_1$ layer separated by two corrugated As$_2$-U layers stacking along *c*-axis and producing a quasi-2D layered structure. The coordination of As atoms around a U atom is shown in the right lower panel of Fig. 1**a**. The U ion resides in a ninefold coordinated environment that has three different U-As bond lengths [20]. Figure 1**c** shows the single-crystal XRD pattern of a cleaved UAs$_2$ sample. Millimeter size sample image is given in the inset. Only (00*l*) reflections are observed, indicating that the easy cleaved plane is the *ab* plane. The calculated lattice constant *c* from the XRD data is 8.125 Å, which is consistent with previous reports [39]. In such a tetragonal lattice with the *c/a* value exceeds 2, the nearest neighbor U-U distance (3.8 Å) is much larger than the so-called Hill limit (~3.5 Å) [14]. The reduction in hybridization with increasing U-U separation would lead to the localization of 5*f*-band electrons. Thus, we may expect that the 5*f*-band electrons of U are mainly localized and form magnetic order at a finite temperature.

Figure 1**d** presents the temperature dependence of magnetic susceptibility *χ(T)* measured with applied magnetic field $\mu_0 H = 1$ T. A clear magnetic transition around 274 K is observed in both zero-field-cooling (ZFC) and field-cooling (FC) modes. The peak observed in *χ(T)* curves at about 274 K may indicate the development of antiferromagnetic correlations. With further cooling down, the magnetization shows a minimum at about 30 K and a Curie-like tail below 10 K. To further clarify the magnetization behavior of UAs$_2$, we performed *M(H)* measurements at various temperatures and show the data in the inset of Fig. 1**d**. We can see that the magnetic moments are proportional to *H* at all temperatures and no saturation is observed below 7 T. The temperature-dependent resistivity of UAs$_2$ at ambient pressure shows a good metallic behavior as illustrated in Fig. 1**e**. In high temperature region, a clear kink feature is observed around $T_N$ = 274 K and the resistivity starts to decrease strongly below $T_N$ due to the formation of the antiferromagnetic order. The low temperature resistivity data can be well described by the typical formula $\rho(T) = \rho_0 + AT^n$ with $n \sim 2.7$. The relatively large *n* value compared to the ideal Fermi-liquid behavior (*n* = 2) may indicate that the joint effect of electron-phonon scattering and magnetic interactions, in consistent with other U-based compounds [40, 41].

## 2. Pressure-induced unconventional superconductivity and upper critical fields

High pressure is an efficient technique to tune the ground state of materials. By applying pressure, the crystal structure and electronic properties can be effectively changed, leading to the appearance of emergent physics, such as unconventional superconductivity and quantum criticality, *etc.* [42, 43, 44] Especially in heavy fermion systems, pressure can easily tune the magnetic order and show rich and unexpected phase diagrams [11, 17-19, 45]. Therefore, we performed *in situ* high-pressure electrical resistance measurements by using a BeCu-type diamond pressure cell (DAC) apparatus. We present the temperature-dependent resistance *R(T)* curves for sample 1 with pressures up to 40.5 GPa in Fig. 2**a**. At a first glance, we find that the room temperature resistance initially remains basically unchanged with increasing pressure, then suddenly drops above a critical pressure ($P_c \sim 20$ GPa), and finally slowly decreases with the further increase of pressure as illustrated in Fig. 2**b**. The inset of Fig. 2**b** shows a photo for the sample configuration in the DAC under high pressures. A conventional four-probe van der Pauw method is adopted in our high-pressure resistive measurements. Previous studies on the high-pressure crystal structure for UAs$_2$ revealed a structure phase transition from tetragonal to orthorhombic in the temperature range about 15-18

GPa [37, 38]. Therefore, we may attribute the sudden change of resistance at room temperature, as shown by the grey region in Fig. 2**b,** to the pressure-induced structural phase transition in UAs$_2$.

We now turn to the fine structures of *R(T)* curves under different pressures. With increasing pressure, the $T_N$ gradually reduces to a lower temperature, and the AFM transition becomes blurred as evidenced from the *R(T)* curves in the upper part of Fig. 2**a**. However, the AFM transition temperature can still be determined by taking the derivative of resistance versus temperature. For $P \geq 22.5$ GPa, there is no sudden drop of resistance at high temperatures and no any anomaly can be discerned in the *R(T)* curves, indicating that the long-range AFM order is completely suppressed due to the structure phase transition. Upon further increasing pressure, the overall shape of *R(T)* curves gradually changes from a convex shape to a concave one, which may indicate that the transport behavior of the title compound gradually changes from non-Fermi-liquid to Fermi liquid behavior under pressure (see the reasoning below).

Accompanied by the vanishing of long-range AFM order, a sudden drop of resistance was observed at low temperatures, as depicted in Fig. 2**c;** zero resistance was observed at about 3 K under 24.4 GPa, indicating that this is a superconducting transition. To show the evolution of the superconducting transition more clearly, we performed high-pressure measurements on another sample (S2) in the similar pressure region where the superconductivity occurs. As shown in Fig. 2**d**, the *R(T)* curve starts to drop at about 22.9 GPa with the $T_c^{onset}$ of about 3.4 K, zero resistance is achieved below about 2.3 K under the pressure of about 24.8 GPa. After that, the superconducting transition temperature gradually increases with increasing pressure and reaches its optimum value at 26.8 GPa. The maximum value of $T_c^{onset}$ and $T_c^{zero}$ are 4.05 K and 3.27 K, respectively. With further increasing pressure, the $T_c$ value is monotonically suppressed and the superconducting transition completely disappears above 31.9 GPa. Thus the superconducting region seems to be quite narrow. Note that the pressure-induced superconductivity and phase transitions are reversible in the present material, in the decompression process we observed the reappearance of superconductivity and the resistivity anomaly corresponding to the AFM state. (Extended Data Fig. 1).

Furthermore, it is worth to note that above the superconducting transitions, the electrical resistance exhibits a linear temperature dependence, as shown by the red line in Fig. 2c. The $T$-linear resistance behavior was widely observed in different material systems with unconventional superconductivity [29-35]. Usually, the normal state $T$-linear resistance, also called as strange metal behavior, is observed in optimal doping/pressure regions for superconductivity and it was thought to be closely related to quantum criticality. In our present experiment, we also find that as the superconducting transition temperature is optimized, the temperature range for the $T$-linear resistance expands and the feature becomes more robust (see Extended Data Fig. 2), indicating the presence of strange-metal state and a possible unconventional pairing mechanism in $UAs_2$ under high pressures.

To further characterize the superconducting state and possible unconventional pairing mechanism in $UAs_2$ under high pressure, we measured the low-temperature $R(T)$ curves under different magnetic fields and present the normalized data in Fig. 3a. As we expect, the superconducting transition shifts to lower temperatures gradually with increasing magnetic fields and finally disappears above 1.9 K with the applied magnetic fields of about 8 T. Considering the relatively weak suppression of superconductivity by the magnetic field, the superconductivity in $UAs_2$ should exist at a much higher magnetic field at zero temperature. Remarkably, when the superconducting transition is completely suppressed at $\mu_0H$ = 9 T, the $R(T)$ curve still exhibits very nice $T$-linear behavior down to 1.9 K, the lowest measured temperature in our present study. This strongly indicates that the strange-metal behavior is the ground state of the title material with optimized superconductivity under high pressures.

Figure 3b displays the derived upper critical fields $\mu_0H_{c2}(T)$ versus temperature for S2 at 25.9 GPa. The $T_c$ is determined as the temperature when the resistance drops to 90% of the normal-state value. As we can see, the $\mu_0H_{c2}$-$T$ plots can be well described by the Ginzburg-Landau (*G-L*) equation,

$$\mu_0H_{c2}(T) = \mu_0H_{c2}(0)[1 - (T/T_c)^2]/[1 + (T/T_c)^2].$$

The obtained zero-temperature upper critical field $\mu_0H_{c2}(0)$ is 11.7 T. And the corresponding coherence length of 5.3 nm is deduced by using $\mu_0 H_{c2}(0) = \Phi_0/2\pi\xi_{GL}^2$, where $\Phi_0 =$ h/2e is the flux quantum. We also conducted the upper critical field measurements on different pressures and samples, see Extended Data Fig. 3. The estimated values of $\mu_0H_{c2}(0)$ in different runs of measurements are quite close to each other, which is about 12 T. More importantly, the value of $\mu_0H_{c2}(0)$ strongly exceeds the corresponding Pauli paramagnetic limit field ($\mu_0H_p$ = 1.84$T_c$ = 6.62 T for $T_c$ = 3.6 K). This naturally puts the superconductivity in UAs$_2$ in the same category as that in UTe$_2$, manifesting an exotic pairing mechanism in UAs$_2$ under high pressures.

## 3. *T-P* phase diagram of UAs$_2$

Based on above results, we construct a temperature-pressure phase diagram for UAs$_2$, as depicted in Fig. 4. At ambient pressure, UAs$_2$ crystallizes in a typical anti-Cu$_2$Sb type structure and orders antiferromagnetically below 274 K. With increasing pressure, $T_N$ shows a monotonic decrease down to below 100 K and suddenly collapses completely at a critical pressure $P_c$ of about 20 GPa. Note that the values of $T_N$ are extracted from the peak position of the d$R$/d$T$ curves (see Extended Data Fig. 4). Above $P_c$, superconductivity emerges along with a possible structure phase transition from tetragonal to orthorhombic symmetry [37, 38]. With further increasing pressure, superconductivity gets enhanced first and then becomes gradually weakened, finally vanishes above about 32 GPa, leaving a dome-like superconducting region.

In order to get a deep insight about the transport behavior of UAs$_2$ under high pressures, we analyze the normal state resistance by fitting the *R(T)* curve with the formula $R(T) = R_0 + AT^n$, where $R_0$ represents the residual resistance, *A* and *n* are the prefactor and exponent of the power-law function. The detailed fitting curves under different pressures and the evolution of *A* and *n* with pressure for S1 are given in Extended Data Figs. 5 and Fig. 6. Similar data for A and n as well as the phase diagram were also found for S2, the only difference is the pressure for optimal superconductivity is slightly higher, see Extended Data Fig. 7. We also show the pressure-dependent exponent *n* in a color plot in the lower part of Fig. 4. At ambient and low pressures below about 4 GPa, *n* is larger than 2. After that, the low-temperature transport behavior changes to a typical Fermi liquid behavior with *n* close to 2. Note that a strange downward curvature of resistance versus temperature was observed in the pressure region from 5 to 20 GPa in the low temperature limit. We don't know yet what is the reason for this downward shape in the low pressure region, thus we mark it as "bad metal" in Fig.4. Just at the threshold of pressure around $P_c \approx$ 20 GPa, the *n* value changes abruptly from around 2 (for a Fermi liquid) to 1 corresponding to a strange metal behavior. It seems that the *T*-linear behavior maintains when superconductivity is present at low temperatures. In this critical pressure region, when superconductivity has the highest transition temperature, the *T*-linear behavior extends to the widest temperature region, see Extended Data Fig. 2. The *n* value gradually increases to 2 after the pressure surpasses about 30 GPa, and accordingly the superconductivity is gone. Furthermore, according to the literatures [31, 43, 47], the $T_c$ and quantum criticality often correlate with the prefactor *A*, producing a peak at the optimal doping or pressure. This happens also to our sample here, as shown in Fig. 4 and Extended Data Fig. 6 for S1; Extended Data Fig.7 for S2. The systematic evolution of resistivity exponent *n* and prefactor *A*, together with the emergence of superconductivity gives strong evidence for a QCP around $P_c$. The observed *T*-linear dependence of resistance (*n* ~ 1) around $P_c$ in UAs$_2$ on approaching the optimal superconducting region implies a close relationship between the strange metal behavior and the emergence of superconductivity.

## Discussion and perspectives

As discussed above, our results give a strong indication that the superconductivity observed in UAs$_2$ under high pressures is unconventional. From the *T-P* phase diagram, we can see that the superconductivity emerges around the border of antiferromagnetic order, which resembles those of many unconventional superconductors [46, 47, 48]. In those cases, the pairing mechanism was widely perceived to link with the magnetism. Moreover, the concurrence of the strange-metal behavior and optimal superconductivity also reflects the unconventional nature of the title compounds. Furthermore, the superconductivity in UAs$_2$ is quite robust against magnetic fields. The unusual large

$\mu_0H_{c2}(0)$ exceeding the Pauli limit provides another strong evidence for unconventional superconductivity, perhaps spin-triplet in nature, resembling its siblings, i.e., UTe$_2$ [1,2,3]. In this sense, we might expect the reentry of superconductivity at lower temperatures or the FFLO states when a higher magnetic field is applied [1,2, 12,49]. It is highly desired to detect the pairing symmetry in UAs$_2$ under a pressure by using the nuclear magnetic resonance (NMR) or muon spin relaxation (μSR).

This novel pairing state occurs most likely because the magnetic moment of uranium atoms order themselves in a ferromagnetic way within the plane. Although the superconductivity was discovered in UAs$_2$ under a pressure, it will certainly inspire further efforts on the interesting superconducting state in the mysterious uranium-based compounds. First, the maximum transition temperature $T_c$ reaches 4 K which is doubled compared with UTe$_2$, this suggests that the $T_c$ can be improved higher in related systems with proper modifications to the electronic structure, for example by using chemical doping. Second, the present study reveals a phase diagram with many interesting effects merged together, such as antiferromagnetic order, robust superconductivity against magnetic field, strange metal state, quantum criticality, etc. This was not seen in UTe$_2$ at ambient pressure. Finally, we would like to point out that the strong electron correlation effect originating from the dual natures, *i.e*, itineracy and localization, of *f*-band electrons from uranium may play an important role in the emergence of the quantum phase transitions observed in UAs$_2$ under high pressures [50]. Our work not only adds a new member to the interesting uranium-based system with the achievement of superconductivity at a much higher transition temperature, but also shows a close connection with the strange metal behavior. Therefore, we may expect to explore more unconventional superconductors with possible spin-triplet paring in compounds containing both partially filled *f-band* and *d-band* electrons, like uranium in UTe$_2$ and UAs$_2$.

## Conclusion

In summary, we have studied the electric transport properties of UAs$_2$ under pressures and discovered the pressure-induced unconventional superconductivity with the optimal $T_c$ of about 4 K, which is the highest value among the uranium-based compounds with 5*f*-band electrons. The $\mu_0 H_{c2}(0)$ is found to exceed the Pauli paramagnetic limit. In addition, we found that the superconductivity emerges in the vicinity of antiferromagnetic order, which is accompanied by a structure phase transition. More importantly, we found the concurrence of superconductivity and the strange-metal behavior, showing a clear feature of quantum criticality. All these interesting discoveries suggest that the superconductivity in $UAs_2$ is unconventional and may host the spin-triplet pairing. Our results will open a new avenue for finding unconventional superconductivity in more uranium-based compounds with 5*f*-band electrons.

## Method

High-quality $UAs_2$ crystals were successfully grown by the chemical vapor transport method. [22, 28] Initially, high-purity U block (3N) and As granules (4N) were mixed in a 1:2.02 mole ratio. This mixture was loaded into a 13-cm-long quartz ampoule with an inner diameter of 20 mm and sealed under a pressure of approximately $10^{-3}$ Pa. Iodine, with a density of 3-5 mg/cm$^3$, served as the transport agent. The sealed ampoule was placed into a tube furnace equipped with two heating zones. The temperature was first raised gradually to (950 °C, 850 °C) over 5 hours and maintained for 48 hours. Subsequently, the temperature profile was adjusted to (750 °C, 950 °C) over 2 hours and maintained for one week to facilitate crystal growth. Upon cooling to room temperature, numerous layer-shaped $UAs_2$ single crystals were successfully obtained within the quartz ampoule.

The crystal structure of $UAs_2$ alloys was identified by x-ray diffraction (XRD) with Cu *Kα* radiation (Bruker; D8 Advance diffractometer; λ = 1.541 Å). The lattice constants are calculated by using the software of TOPAS 4.2 [51]. The crystal structure is made with VESTA software [52]. DC magnetization measurements were performed with a SQUID-VSM-7 T (Quantum Design). Temperature-dependent electrical resistance measurements under both ambient and high pressures were carried out on a physical property measurement system PPMS-9 T (Quantum Design).

Temperature and magnetic field-dependent resistance under high pressures were measured by a four-probe van der Pauw method with platinum foil as electrodes. The samples with an average size of 100×80 μm$^2$ and a thickness of 20 μm were loaded in our DAC setups (DACPPMS-ET225, Shanghai Anvilsource Material Technology Co., Ltd) with T301 steel gasket to generate the pressure above 40 GPa. In our study, we use soft KBr as the pressure-transmitting medium and the pressure values were determined by the ruby fluorescence method [53].

**Acknowledgments**

This work was supported by the National Key R&D Program of China (Grant No. 2022YFA1403201), National Natural Science Foundation of China (Grant Nos. 12061131001, 11927809, 12204231, and 123B2055), the Natural Science Foundation of Hunan Province, China (Grant No. 2023JJ10051).

**Author contributions:** H.H.W. conceived and supervised the whole study. The single crystal samples were grown by B.-B. Z. and C. Z. The sample characterizations at ambient pressure were done by Z.-N.X., and Q.L. The high-pressure electrical transport measurements were performed by Y.-J. Z., Z.-N. X., and Q. L. with assistance from H.-H. W. H.-H. W. and Q. L. analyzed the experimental data and wrote the manuscript with the inputs from all co-authors.

**Competing interests:** The authors declare that they have no competing interests.

**Data and materials availability:** All data needed to evaluate the conclusions in the paper are present in the paper. Additional data related to this paper may be requested from the authors.

**References**

[1] Ran, S. et al. Nearly ferromagnetic spin-triplet superconductivity. *Science* **365**, 684–687 (2019).

[2] Aoki, D. et al. Unconventional Superconductivity in Heavy Fermion UTe$_2$. *J. Phys. Soc. Jpn.* **88**, 043702 (2019).

[3] Knebel, G. et al. Field-Reentrant Superconductivity Close to a Metamagnetic Transition in the Heavy-Fermion Superconductor UTe$_2$. *J. Phys. Soc. Jpn.* **88**, 063707 (2019).

[4] Jiao, L. et al. Chiral superconductivity in heavy-fermion metal UTe$_2$. *Nature* **579**, 523–527 (2020).

[5] Hayes, I. M. et al. Multicomponent superconducting order parameter in UTe$_2$. *Science* **373**, 797–801 (2021).

[6] Sato, M. & Ando, Y. Topological superconductors: a review. *Rep. Prog. Phys.* **80**, 076501 (2017).

[7] Alicea, J. New directions in the pursuit of Majorana fermions in solid state systems. *Rep. Prog. Phys.* **75**, 076501 (2012).

[8] Schemm, E. R., Gannon, W. J., Wishne, C. M., Halperin, W. P. & Kapitulnik, A. Observation of broken time-reversal symmetry in the heavy-fermion superconductor UPt$_3$. *Science* **345**, 190–193 (2014).

[9] Saxena, S. S. et al. Superconductivity on the border of itinerant-electron ferromagnetism in UGe$_2$. *Nature* **406**, 587-592 (2000).

[10] Aoki, D. et al. Coexistence of superconductivity and ferromagnetism in URhGe. *Nature* **413**, 613–616 (2001).

[11] Huy, N. T. et al. Superconductivity on the Border of Weak Itinerant Ferromagnetism in UCoGe. *Phys. Rev. Lett.* **99**, 067006 (2007).

[12] Ran, S. et al. Extreme magnetic field-boosted superconductivity. *Nat. Phys.* **15**, 1250–1254 (2019).

[13] Duan, C. et al. Resonance from antiferromagnetic spin fluctuations for superconductivity in UTe$_2$. *Nature* **600**, 636–640 (2021).

[14] Aoki, D. et al. Unconventional superconductivity in UTe$_2$. *J. Condens. Matter*

*Phys.* **34**, 243002 (2022).

[15] Aishwarya, A. et al. Magnetic-field-sensitive charge density waves in the superconductor UTe$_2$. *Nature* **618**, 928–933 (2023).

[16] Gu, Q. et al. Detection of a pair density wave state in UTe$_2$. *Nature* **618**, 921–927 (2023).

[17] Thomas, S. M. et al. Evidence for a pressure-induced antiferromagnetic quantum critical point in intermediate-valence UTe$_2$. *Sci. Adv.* **6**, eabc8709 (2020).

[18] Li, D. et al. Magnetic Properties under Pressure in Novel Spin-Triplet Superconductor UTe$_2$. *J. Phys. Soc. Jpn.* **90**, 073703 (2021).

[19] Honda, F. et al. Pressure-induced Structural Phase Transition and New Superconducting Phase in UTe$_2$. *J. Phys. Soc. Jpn.* **92**, 044702 (2023).

[20] Amoretti, G., Blaise, A. & Mulak, J. Crystal field interpretation of the magnetic properties of UX$_2$ compounds (X = P, As, Sb, Bi). *J. Magn. Magn. Mater.* **42**, 65–72 (1984).

[21] Blaise, A. et al. Physical properties of uranium dipnictides. *Conference on rare earths and actinides* **37**, 184–189 (1977).

[22] Wiśniewski, P. et al. Cylindrical Fermi surfaces of UAs$_2$ and UP$_2$. *Physica B* **281**, 769–770 (2000).

[23] Troć, R., Leciejewicz, J. & Ciszewski, R. Antiferromagnetic Structure of Uranium Diphosphide. *Phys. Stat. Sol.* **15**, 515–519 (1966).

[24] Leciejewicz, J., Troć, R., Murasik, A. & Zygmunt, A. Neutron-Diffraction Study of Antiferromagnetism in USb$_2$ and UBi$_2$. *Phys. Stat. Sol.* **22**, 517–526 (1967).

[25] Wisniewski, P. et al. Shubnikov-de Haas effect study of cylindrical Fermi surfaces in UAs$_2$. *J. Magn. Magn. Mater.* **12**, 1971–1980 (2000).

[26] Aoki, D. et al. Cylindrical Fermi surfaces formed by a fiat magnetic Brillouin zone in uranium dipnictides. *Philos. Mag. B* **80**, 1517–1544 (2000).

[27] Chen, Q. Y. et al. Orbital-Selective Kondo Entanglement and Antiferromagnetic Order in USb$_2$. *Phys. Rev. Lett.* **123**, 106402 (2019).

[28] Ji, X. et al. Direct observation of coexisting Kondo hybridization and antiferromagnetic state in UAs$_2$. *Phys. Rev. B* **106**, 125120 (2022).

[29] Legros, A. et al. Universal T-linear resistivity and Planckian dissipation in overdoped cuprates. *Nat. Phys.* **15**, 142–147 (2019).

[30] Cai, S. et al. The breakdown of both strange metal and superconducting states at a pressure-induced quantum critical point in iron-pnictide superconductors. *Nat Commun.* **14**, 3116 (2023).

[31] Zhang, Y. et al. High-temperature superconductivity with zero resistance and strange-metal behaviour in $La_3Ni_2O_{7-\delta}$. *Nat. Phys.* 1-5 (2024).

[32] Nguyen, D. H. et al. Superconductivity in an extreme strange metal. *Nat. Commun.* **12**, 4341 (2021).

[33] Doiron-Leyraud, N. et al. Correlation between linear resistivity and T c in the Bechgaard salts and the pnictide superconductor $Ba(Fe_{1-x}Co_x)_2As_2$. *Phys. Rev. B* **80**, 214531 (2009).

[34] Lee, K. et al. Linear-in-temperature resistivity for optimally superconducting $(Nd,Sr)NiO_2$. *Nature* **619**, 288–292 (2023).

[35] Cao, Y. et al. Strange Metal in Magic-Angle Graphene with near Planckian Dissipation. *Phys. Rev. Lett*. **124**, 076801 (2020).

[36] Coleman, P. & Schofield, A. J. Quantum criticality. *Nature* **433**, 226-229 (2005).

[37] Benedict, U. & Holzapfel, W. B. High-pressure studies-Structural aspects. *Handbook on the Physics and Chemistry of Rare Earths* **17**, 245-300 (1993).

[38] Gerward, L., Olsen, J. S., Benedict, U., Dancausse, J. P. & Heathman, S. High-pressure X-ray diffraction studies of $ThS_2$, $US_2$ and other $AnX_2$ and AnXY compounds. *AIP* **309**, 453-456 (1994).

[39] Trzebiatowski, W., Sepichowska, A. & Zygmunt, A. Magnetic properties of uranium compounds with arsenic or antimony. *Bull. Acad. Polon. Sci., Ser. Sci. Chem.* **12,** 687-690 (1964).

[40] Wawryk, R. Magnetic and transport properties of $UBi_2$ and $USb_2$ single crystals. *Philos. Mag.* **86**, 1775–1787 (2006).

[41] Wawryk, R., Henkie, Z., Cichorek, T., Geibel, C. & Steglich, F. Transport Properties of $URhGa_5$ Single Crystals. *Phys. Stat. Sol.* **232**, R4–R6 (2002).

[42] Sun, H. et al. Signatures of superconductivity near 80 K in a nickelate under high

pressure. *Nature* **621**, 493–498 (2023).

[43] Shen, B. et al. Strange-metal behaviour in a pure ferromagnetic Kondo lattice. *Nature* **579**, 51–55 (2020).

[44] Yang, P. T. et al. Pressured-induced superconducting phase with large upper critical field and concomitant enhancement of antiferromagnetic transition in EuTe$_2$. *Nat. Commun.* **13**, 2975 (2022).

[45] Jeffries, J. R., Stillwell, R. L., Weir, S. T., Vohra, Y. K. & Butch, N. P. Emergent ferromagnetism and T-linear scattering in USb$_2$ at high pressure. *Phys. Rev. B* **93**, 184406 (2016).

[46] Chen, H. et al. Pressure induced superconductivity in the antiferromagnetic Dirac material BaMnBi$_2$. *Sci Rep.* **7**, 1634 (2017).

[47] Liu, Z. Y. et al. Pressure-Induced Superconductivity up to 9 K in the Quasi-One-Dimensional KMn$_6$Bi$_5$. *Phys. Rev. Lett.* **128**, 187001 (2022).

[48] Honda, F. et al. Pressure-induced superconductivity and large upper critical field in the noncentrosymmetric antiferromagnet CeIrGe3. *Phys. Rev. B* **81**, 140507 (2010).

[49] Kasahara, S. et al. Evidence for an Fulde-Ferrell-Larkin-Ovchinnikov State with Segmented Vortices in the BCS-BEC-Crossover Superconductor FeSe. *Phys. Rev. Lett.* **124**, 107001 (2020).

[50] Sheng, Q. G., Cooper, B. R. & Lim, S. P. Trend of f-electron localization and itinerancy in rare-earth and light-actinide systems. *J. Appl. Phys*. **73**, 5409–5411 (1993).

[51] Rietveld, H. M. A profile refinement method for nuclear and magnetic structures. *J. Appl. Crystallogr.* **2**, 65–71 (1969).

[52] Momma, K. & Izumi, F. VESTA : a three-dimensional visualization system for electronic and structural analysis. *J. Appl. Crystallogr.* **41**, 653–658 (2008).

[53] Mao, H. K., Xu, J. & Bell, P. M. Calibration of the ruby pressure gauge to 800 kbar under quasi-hydrostatic conditions. *J. Geophys. Res.* **91**, 4673–4676 (1986).

# Figures and legends

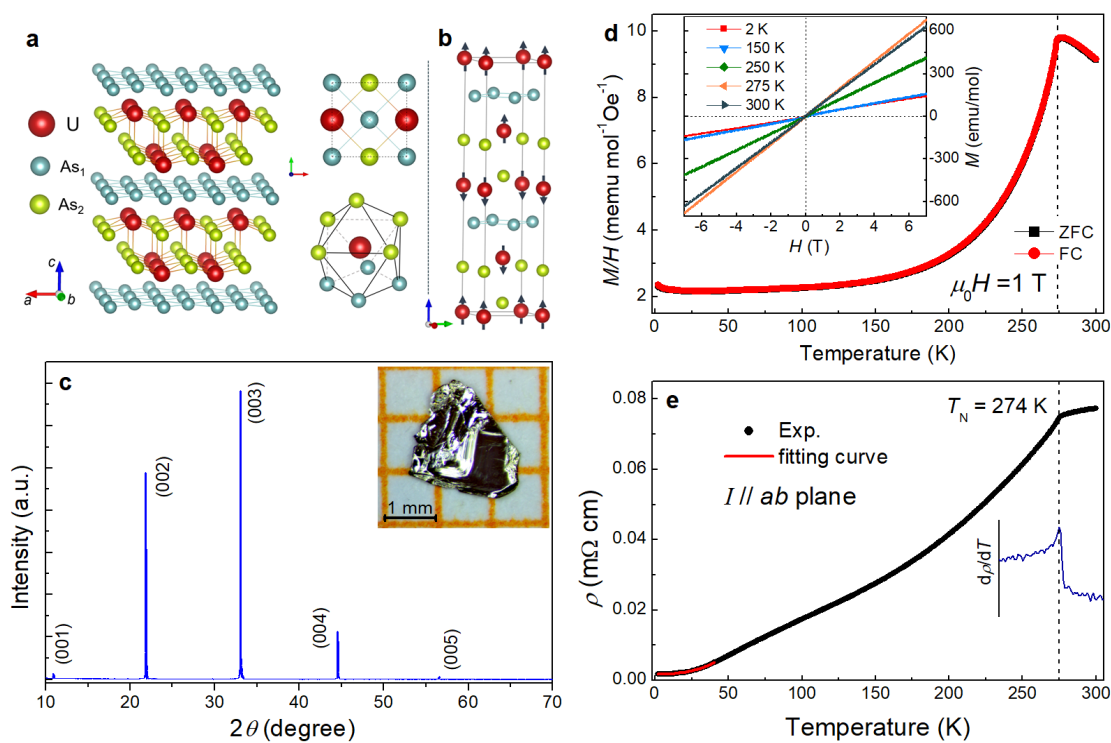

**Fig.1| Structure, magnetic susceptibility and resistivity at ambient pressure. a** Crystal structure of tetragonal UAs$_2$. The red, cyan, and yellow spheres represent the U and As atoms, respectively. Right upper panel: Top view of the structure, where the As$_1$ atoms form a flat square lattice. Right lower panel: Ninefold coordination of U in UAs$_2$. **b** The magnetic structure of tetragonal UAs$_2$. **c** Single-crystal x-ray diffraction pattern of UAs$_2$. The inset shows a typical photograph of the as-grown single crystal. **d** Temperature-dependent susceptibility of UAs$_2$ with magnetic field $\mu_0 H = 1$ T along $c$-axis. Inset shows the magnetization hysteresis $M(H)$ loops at various temperatures. **e** Temperature-dependent resistivity of UAs$_2$ at ambient pressure. The electric current is parallel to the $ab$ plane. The red line in low temperature region is the fitted result by using the equation $\rho(T) = \rho_0 + AT^n$, where $\rho_0$ is the residual resistivity, $n$ and $A$ are the fitting parameters.

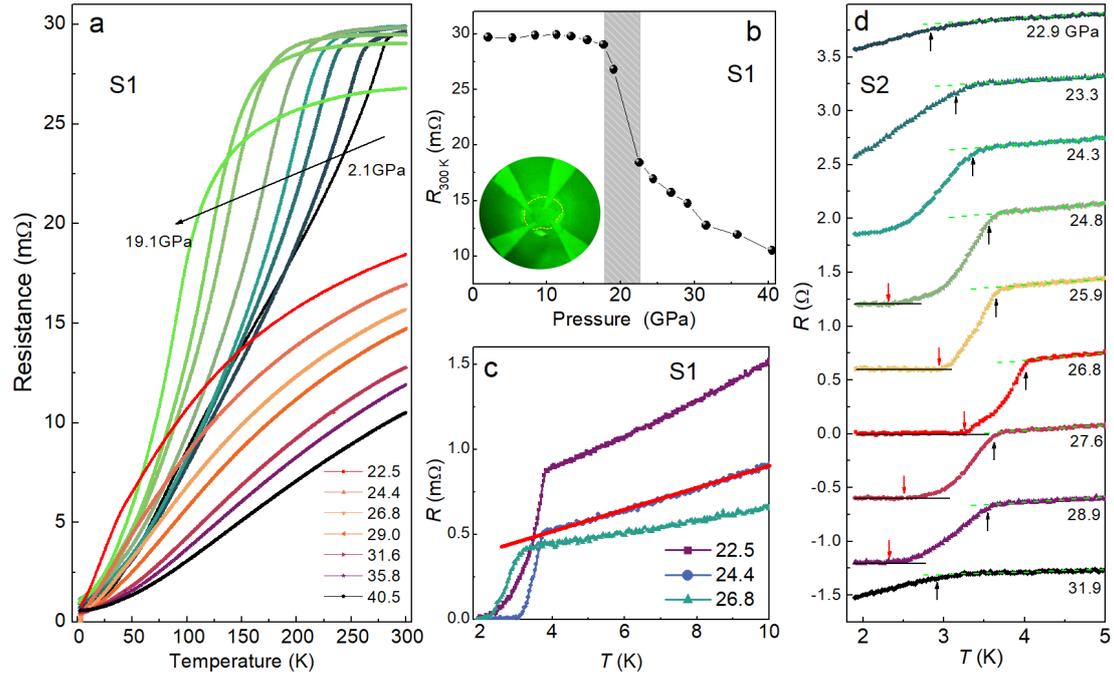

**Fig.2 | Temperature dependence of resistance under pressures. a** Temperature-dependent resistance $R(T)$ of UAs$_2$ for sample 1 (S1) under various pressures up to 40.5 GPa. **b** Pressure-dependent resistance at 300 K. Inset shows the sample and electrode configuration inside the DAC apparatus, and the dashed line gives roughly the outline of the sample surrounded by the pressure transmitting material KBr. **c** Enlarged view of $R(T)$ curves in low-temperature region at selected pressures. $T$-linear dependence of resistance is observed in the normal states above the superconducting transition. **d** The evolution of superconducting transition under different pressures for another sample (S2). The black and red arrows represent the $T_c^{90\%}$ and $T_c^{zero}$, respectively. The dotted green lines give the extrapolation of the normal state resistance. Except for data at $P$ = 26.8 GPa, all the curves were shifted vertically for clarity.

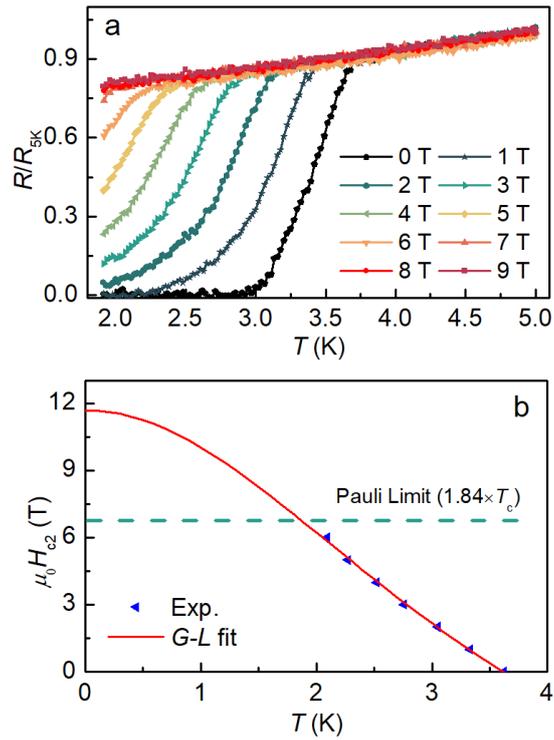

**Fig.3 | Superconducting transitions under different magnetic fields and upper critical field. a** Normalized resistance for S2 at 25.9 GPa under various magnetic fields up to 9 T. **b** Temperature-dependence of the upper critical field $\mu_0 H_{c2}$ for UAs$_2$ under the pressure of 25.9 GPa. The red line represents the fitting curve by using the Ginzburg–Landau (*G–L*) equation and the corresponding Pauli-limiting fields $\mu_0 H_p(0)$ are given by the horizontal green dashed line.

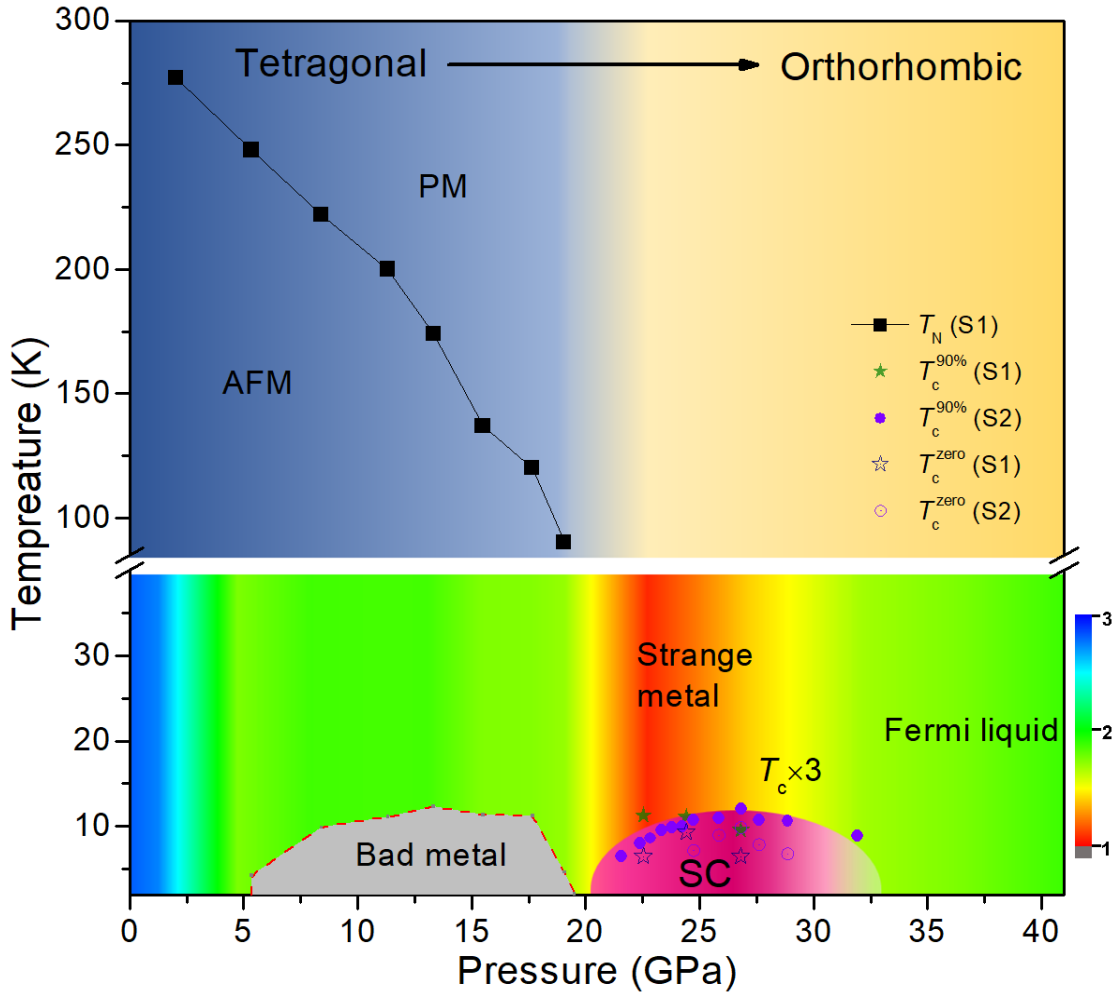

**Fig.4 | *T-P* Phase diagram.** Phase diagram of AFM order, superconductivity, and strange-metal behavior in UAs$_2$ under high pressures. Upper panel: The solid circles display the pressure-dependent $T_N$ determined from the *R(T)* data. PM and AFM denote paramagnetic and antiferromagnetic ordered region. Different colors represent the change of crystal structure symmetry from tetragonal to orthorhombic. Lower panel: *T-P* phase diagram of UAs$_2$ in low-temperature region. The color indicates the fitted exponent *n* by using the formula, $R(T) = R_0 + AT^n$, for S1 within the low-temperature region from 2 to 40 K. Bad metal, strange metal and Fermi liquid represent the regions where the exponent *n* is less than 1, close to 1, and close to 2, respectively. SC represents the superconducting region, where the value of $T_c$ is multiplied by a factor of three. Here, we define $T_c$ as the temperature with 90% normal state resistance as indicated by the arrows in Fig. 2**d**. In total we have carried out 4 runs of experiment, and all of them show superconductivity and *T*-linear dependence of resistance in the normal state when the sample is under a certain pressure. Thus, the effect reported are easily reproduced.

# Extended Data Figures

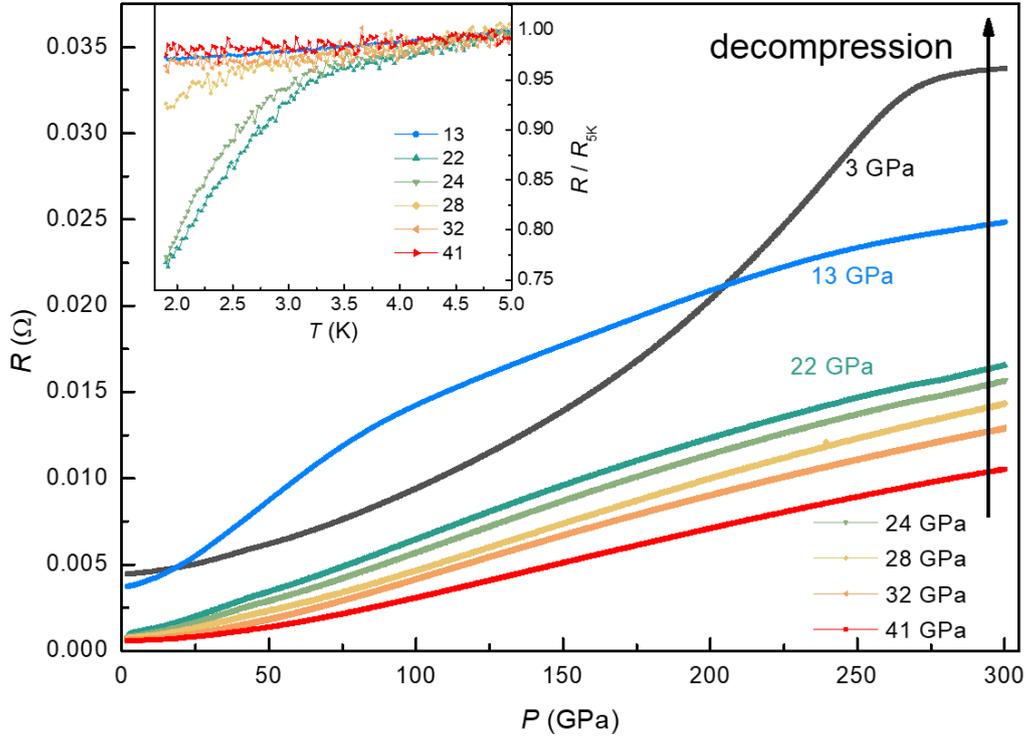

**Extended Data Fig. 1 | Resistance in the decompression process.** Temperature-dependent resistance $R(T)$ of UAs$_2$ for sample 1 (S1) in the decompression process. The inset shows the normalized pressure-temperature dependence of resistance ($R / R_{5K}$ - $T$) of UAs$_2$ from 1.9 to 5 K. As the pressure is gradually removed, superconductivity reappears, and then gradually disappears again until it enters an antiferromagnetic state. Note that during the decompression process, the superconductivity cannot enter a zero-resistance state, which may be because the superconductivity in UAs$_2$ is particularly sensitive to the pressure environment. Such behavior is very similar to that of many heavy fermion superconductors.

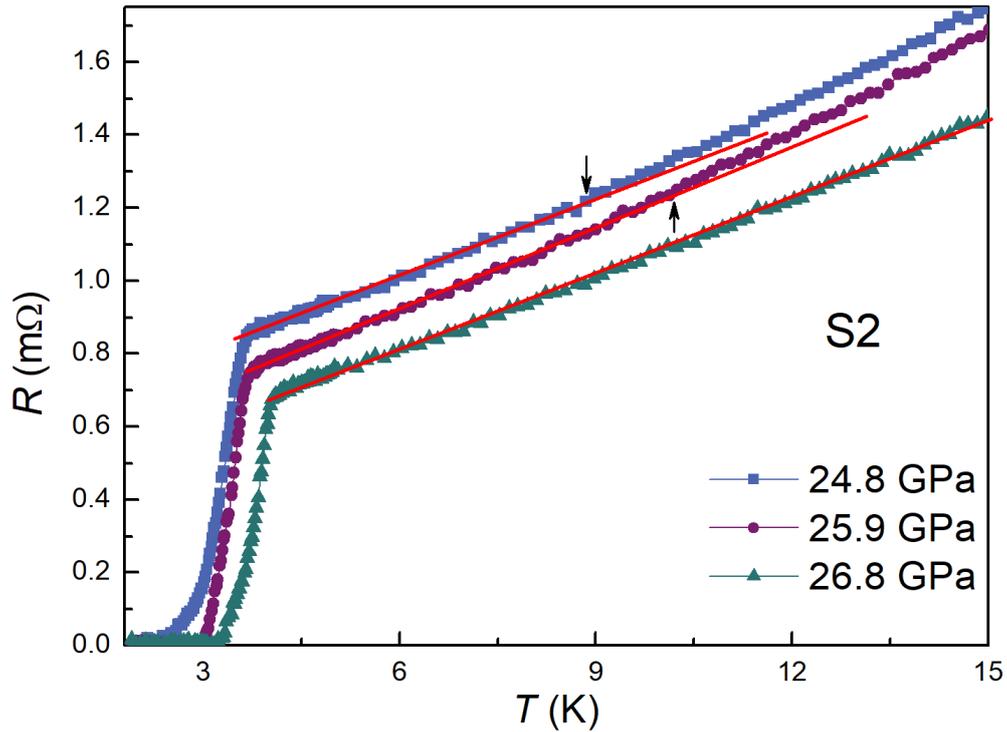

**Extended Data Fig. 2 | Superconducting transitions under different pressures.** Enlarged view of *R(T)* curves in low-temperature region at selected pressures for S2. *T*-linear dependence of resistance is also observed in the normal state in the pressure region with optimal superconductivity. We can see that there is a certain correlation between the superconducting transition $T_c$ and the temperature region for *T*-linear resistance. As the $T_c$ increases, the region for *T*-linear resistance in its normal state becomes wider, showing a feature of quantum critical point around the maximum $T_c$.

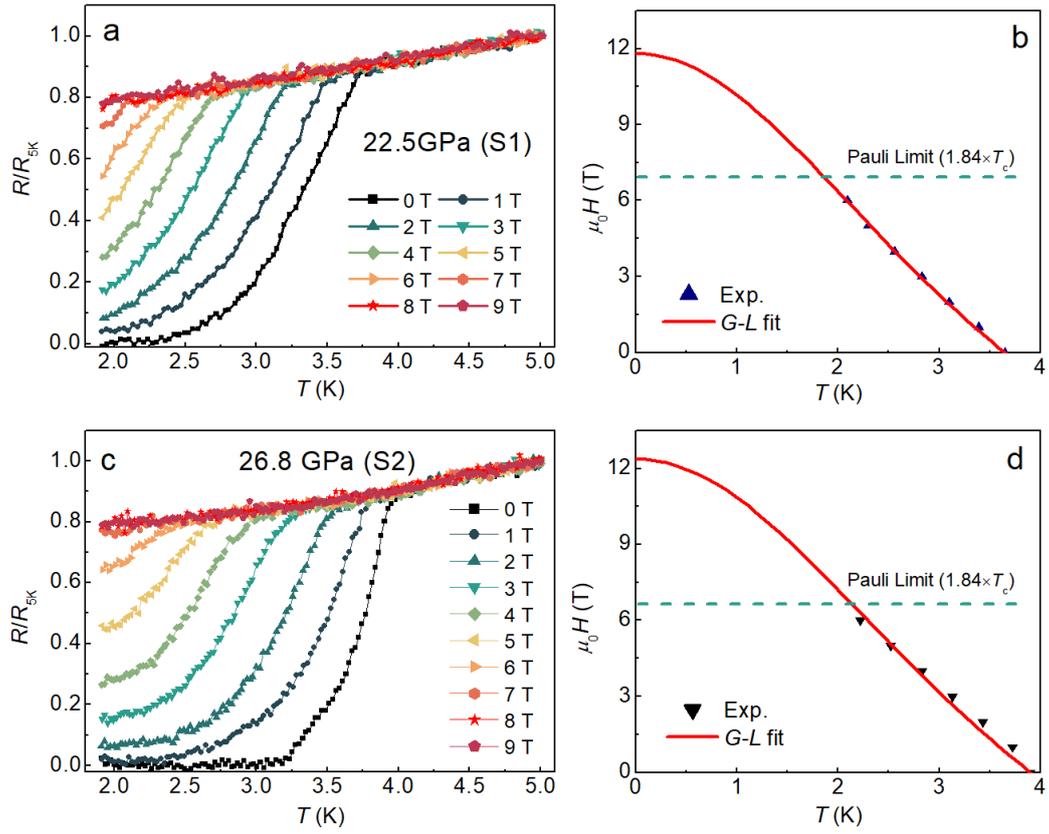

**Extended Data Fig. 3 | Superconducting transitions under different magnetic fields.**
**a** Temperature-dependent resistance (normalized at 5 K) for sample 1 (S1) at 22.5 GPa under various magnetic fields up to 9 T. **c** Temperature-dependent resistance (normalized at 5 K) for sample 2 (S2) at 26.8 GPa under various magnetic fields up to 9 T. **b, d** Temperature-dependence of upper critical field $\mu_0H_{c2}$ for different samples of UAs$_2$ under the pressures shown in the figures. The red lines represent the fitting curves by using the Ginzburg-Landau (*G-L*) equation and the corresponding Pauli-limiting fields $\mu_0H_p(0)$ are given by the horizontal green dashed lines.

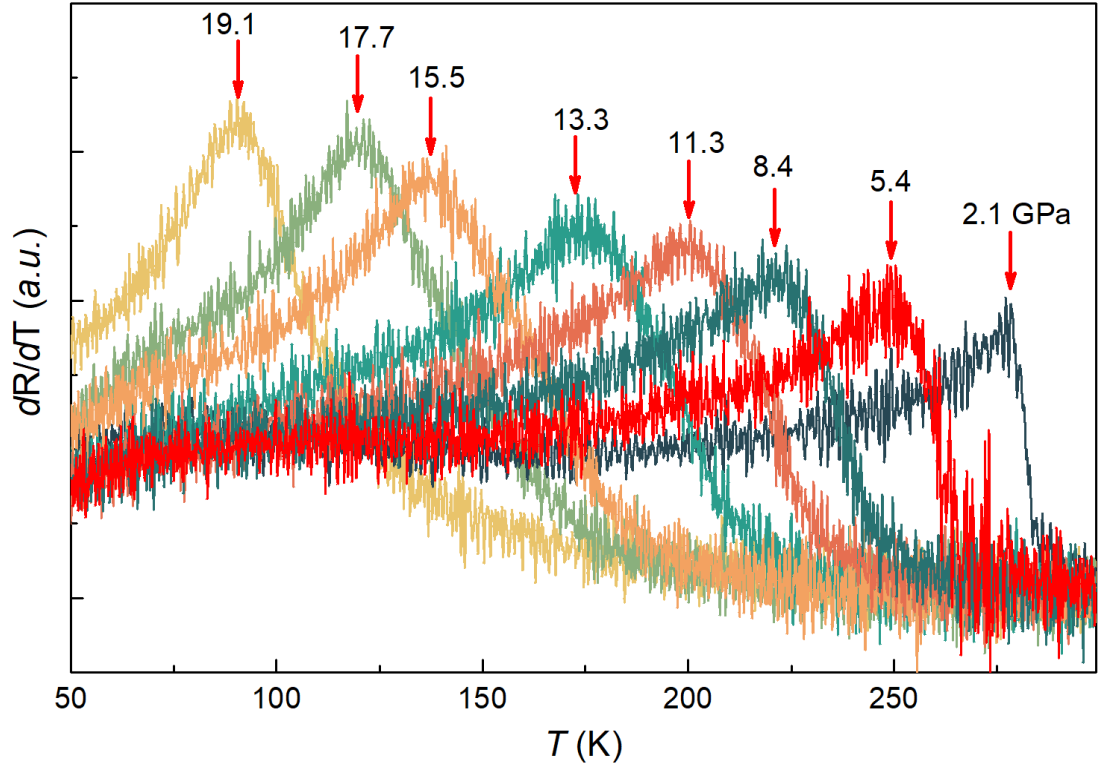

**Extended Data Fig. 4 | Derivative of resistance under different pressures.** The derivative d$R$/d$T$ curve for UAs$_2$ (S1) in the pressure range from 2.1 to 19.1 GPa. The value of $T_N$ is defined as the temperature where the maximum value is observed.

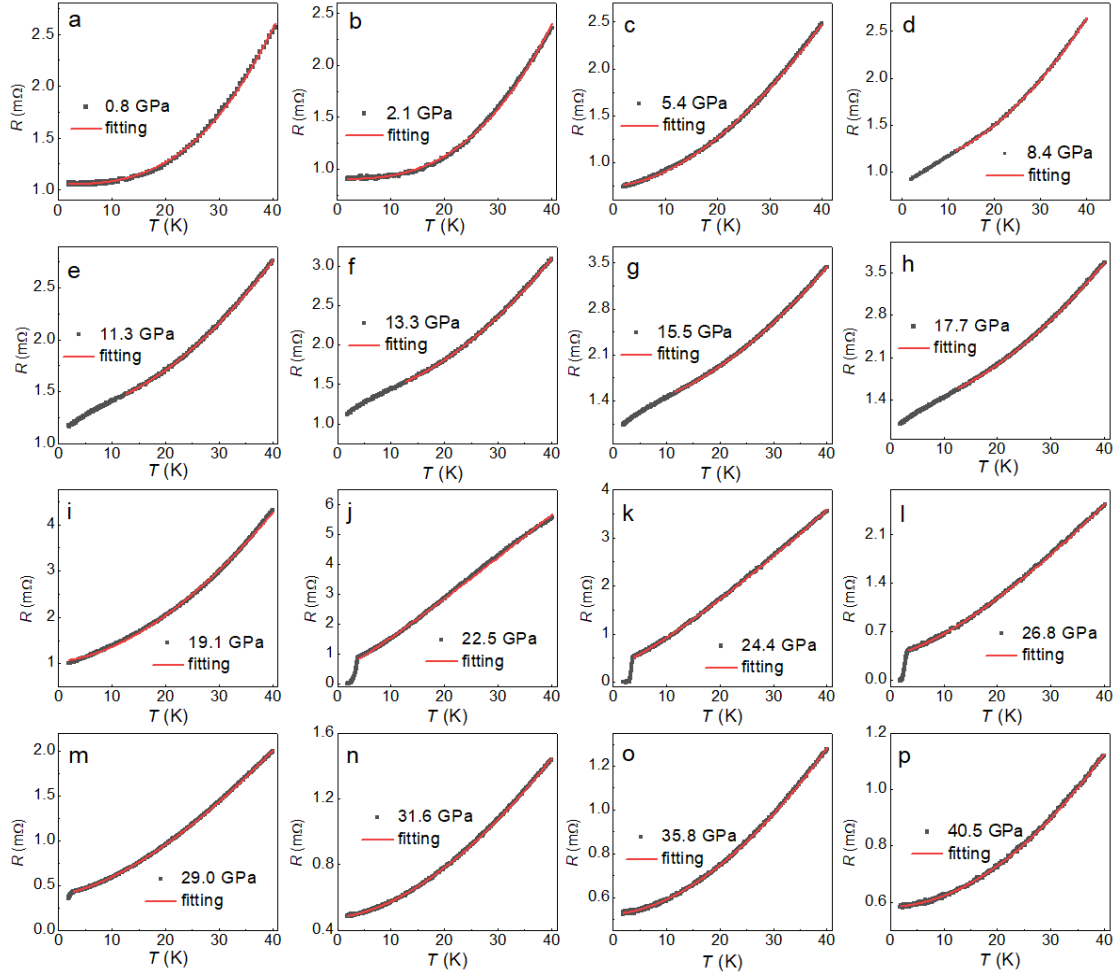

**Extended Data Fig. 5 | Fitting to the low temperature resistance below 40 K. a-p** Temperature-dependent resistance $R(T)$ curves from 2 to 40 K for S1 under various pressures up to 40.5 GPa for S1. The solid red lines are the fitting curves given by the formula $R(T) = R_0 + AT^n$ With fitting up to 40 K. At low pressures below 5.4 GPa, the $R(T)$ curves can be well fitted in the whole temperature region we plotted. At higher pressures, we see a region with negative curvature ($n < 1$) of $R(T)$ curves at low temperatures for the pressures ranging from 8.4 to 19.1 GPa. We termed the transport property in this region as bad metal behavior. Above $P_c \approx 20$ GPa, superconductivity emerges along with the linear-T dependence of resistivity in the normal state. The transport property above $T_c$ can be fitted by above formula, yielding the fitting parameters $A$ and $n$.

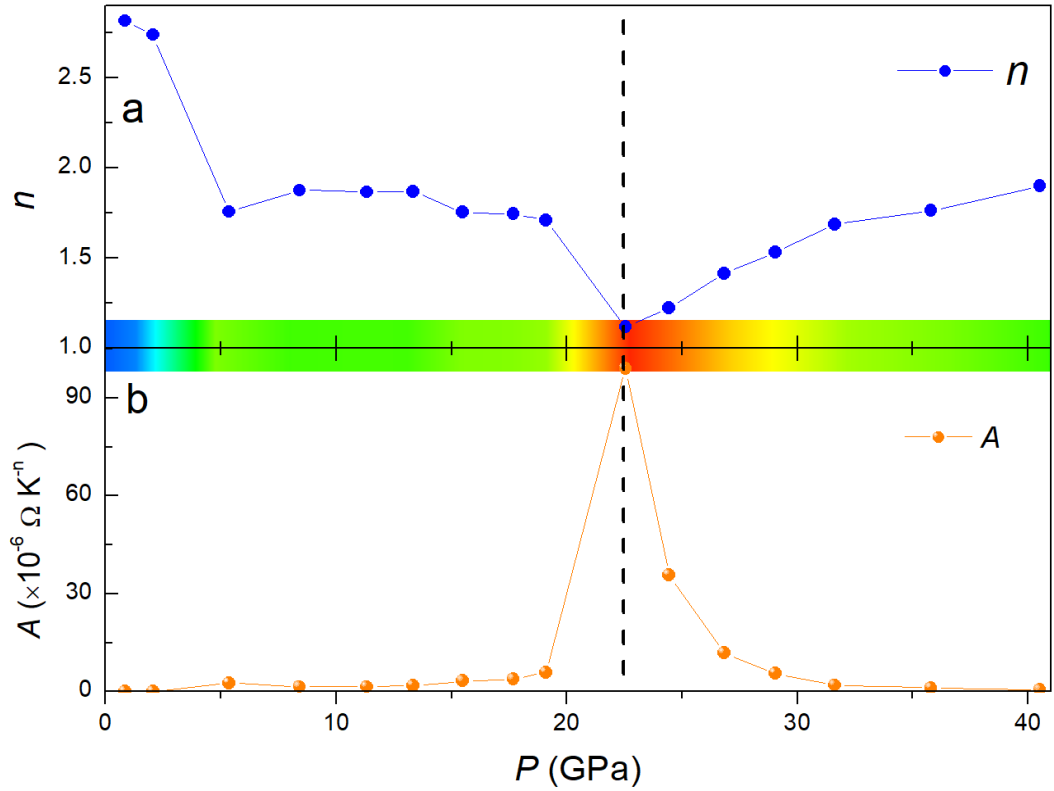

**Extended Data Fig. 6 | The resultant fitting parameter n and A of S1. a, b** Pressure dependence of the exponent $n$ and prefactor $A$ derived from the fitting to the formula $R(T) = R_0 + AT^n$ for sample S1. The middle color plot represents the evolution of the fitted exponent $n$ at different pressures and the black dashed line indicates the possible existence of quantum criticality in $UAs_2$ under high pressure, the color bar is the same as that in Fig. 4. The prefactor $A$ shows a pronounced maximum near the QCP.

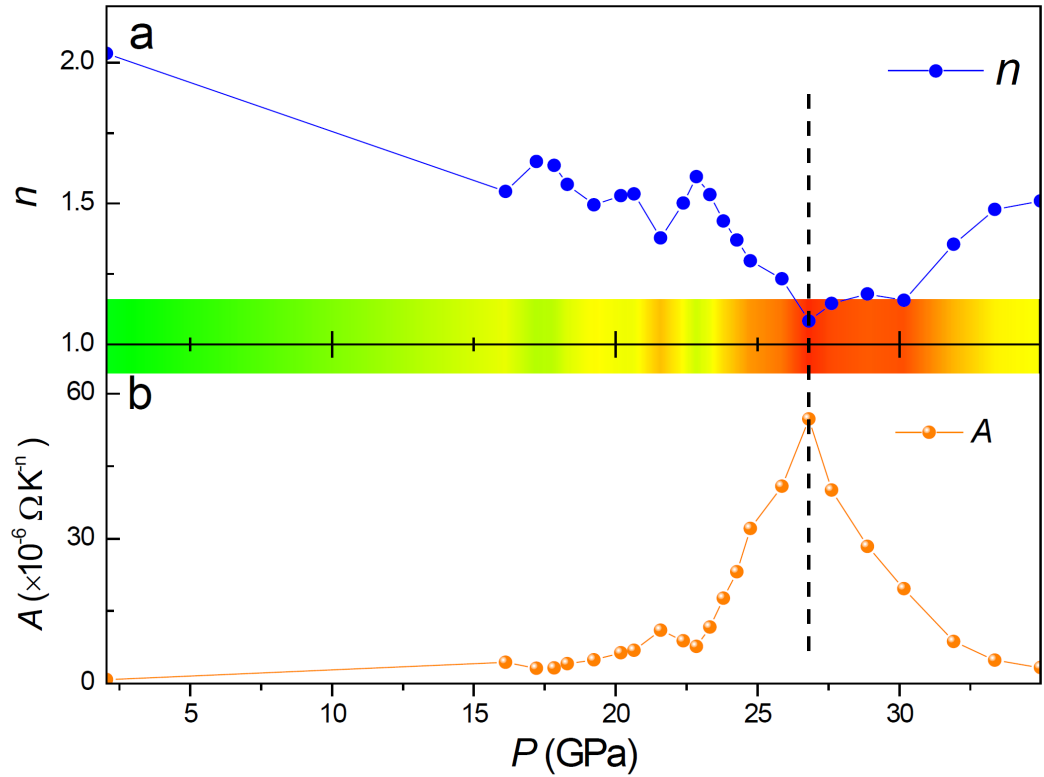

**Extended Data Fig. 7 | The resultant fitting parameter n and A of S2. a, b** Pressure dependence of the exponent $n$ and prefactor $A$ derived from the fitting to the formula $R(T) = R_0 + AT^n$ for sample S2. The middle color plot represents the evolution of the fitted exponent $n$ at different pressures and the black dashed line indicates the possible existence of quantum criticality in $UAs_2$ under high pressure, the color bar is the same as that in Fig. 4. The prefactor $A$ shows a pronounced maximum near the QCP.